\documentstyle[12pt]{article}
\begin{document}
\thispagestyle{empty}

\begin{center}
{\large \bf Dependence of Deep Inelastic Structure \\
Functions on Quark Masses} \\
\vspace*{1cm} 
A.V. Kisselev, V.A. Petrov \\
\vspace*{.5cm}
{\small Institute for High Energy Physics \\
142284 Protvino, Russia} \\
\vspace*{2cm}  
{\bf Abstract} 
\end{center}

\noindent
\begin{quote}
We argue that the difference between the structure functions corresponding
to deep inelastic scattering with and without heavy quarks in the current
fragmentation region scales at high $Q^2$ and fixed (low) $x_{Bj}$.
The lower bound on charm contribution to the total structure function, 
$F_2^c(Q^2,x)$, is calculated and compared with the recent data on 
$F_2^c(Q^2,x)$ from H1 Collaboration.
\end{quote}
\vfill \eject

\section{Introduction}

Quite often mass effects in high--energy collisions are considered as some not 
very spectacular corrections which finally die off. Nonetheless, it appears 
that in $e^+e^-$ annihilation even such overall characteristics as hadron
multiplicities are quite sensitive to the value of masses of the primary
$q \bar q$ pairs~\cite{Data}. 

Recent considerations have shown that 
calculations based on QCD agree well with the data at high enough 
energy~\cite{Petrov-95} and that they yield an asymptotically constant 
difference between multiplicities of hadrons induced by the primary quarks of 
different masses.

In this paper we study a similar effect in a deeply
inelastic process~\cite{Kisselev-95}, \cite{Kisselev-96}. As a by-product,
we estimate heavy quark contributions to the total structure function.

\section{Calculation of quark mass dependence}
Let us consider, for definiteness, deep inelastic scattering of the electron
(muon) off the proton. The hadronic tensor (an imaginary part of the virtual 
photon--proton amplitude) is defined via the electromagnetic current $J_{\mu}$:
\begin{equation}
W_{\mu \nu} (p,q) = \frac{1}{2} (2\pi)^2 \int d^4z \exp (iqz)
< p|[J_{\mu}(z), J_{\nu}(0)]|p >, \label{1}
\end{equation}
where $p$ is the momentum of the proton, $p^2=M^2$, and $q$ is the momentum of
virtual photon, $q^2=-Q^2<0$.

A symmetric part of $W_{\mu \nu}$ has two Lorentz structures:
\begin{equation}
W_{\mu \nu} = \left(- g_{\mu \nu} + \frac{q_{\nu}q_{\nu}}{q^2} 
\right) F_1(Q^2,x) 
+ \frac{1}{pq} \left( p_{\mu} - q_{\mu} \frac{pq}{q^2}\right) 
\left( p_{\nu} - q_{\nu} \frac{pq}{q^2}\right) F_2(Q^2,x), \label{2}
\end{equation}
where the structure functions $F_1$ and $F_2$ depend on $Q^2$ and on the 
variable
\begin{equation}
x = \frac{Q^2}{pq + \sqrt{(pq)^2 + Q^2 M^2}}\raise 2pt \hbox{.} \label{3}
\end{equation}

In what follows we will analyse the structure function $F_2$
of deep inelastic scattering with open charm (beauty) production at small $x$. 
In this section we consider the case of one single quark loop with mass $m_q$
and electric charge $e_q$. A general case will be discussed in Section~3.

At small $x$ a leading contribution to $F_2$ comes from one photon--gluon
fusion subprocess~\cite{Catani-91}:
\begin{equation}
W_{\mu \nu} = \int \frac{d^4k}{(2\pi)^4} \frac{1}{k^4}
C_{\mu \nu}^{\alpha \beta} (q,k;m_q)
 d_{\alpha \alpha'}(k) d_{\beta \beta'}(k)
\Gamma^{\alpha' \beta'} (k,p), \label{4}
\end{equation}
where $k$ is the momentum of the virtual gluon, $k^2<0$. The tensor 
$C_{\mu \nu}^{\alpha \beta}$ denotes an imaginary two gluon irreducible 
part of the photon--gluon amplitude, while $\Gamma^{\alpha \beta}$ describes a 
distribution of the gluon inside the proton. A quantity $d_{\alpha \beta}$
is a tensor part of the gluonic propagator.

Let us choose an infinite momentum frame 
\begin{equation}
p_{\mu} = \biggl(P + \frac{M^2}{4P},0,0, P - \frac{M^2}{4P}\biggr) \raise 2pt
\hbox{.} \label{5}
\end{equation}
Then the gluon distribution 
$\Gamma^{\alpha \beta}$ has to be calculated in the axial gauge $nA=0$ with
a gauge vector $n_{\mu}=(1,0,0,-1)$~\cite{Catani-91}. One can take, 
for instance,
\begin{equation}
n_{\mu} = q_{\mu} + x p_{\mu} \label{6}
\end{equation}
with $x$ defined by Eq.~(\ref{3}).

From Eq.~(\ref{2}) we get
\begin{equation}
\frac{1}{x} F_2 = \biggl[ -g_{\mu \nu} + p_{\mu} p_{\nu}
\frac{3Q^2}{(pq)^2 + Q^2 M^2} \biggl] W^{\mu \nu} \equiv F_2^{(a)} 
+ F_2^{(b)}. \label{7}
\end{equation}
Two terms in the RHS of Eq.~(\ref{7}), $F_2^{(a)}$ and $F_2^{(b)}$, 
correspond to two tensor projectors, $g_{\mu \nu}$ and $p_{\mu}p_{\nu}$.

Note that the structure function $F_L=F_2-2xF_1$ is completely defined 
by the term  $p_{\mu}p_{\nu}$ and, thus, proportional to $F_2^{(b)}$.

By definition, the gluon distribution $\Gamma^{\alpha \beta}$ can be rewritten
in the form
\begin{equation}
\Gamma^{\alpha \beta} = \frac{1}{4\pi} \sum_n \delta ( p + k - p_n) 
< p|I_{\alpha}^g(0)|n >< n|I_{\beta}^g(0)|p >, 
\label{8}
\end{equation}
where $I_{\alpha}^g$ is the conserved current. Both $|p >$ and 
$|n >$ are on shell states that result in
\begin{equation}
k^{\alpha} \Gamma_{\alpha \beta} = 0. \label{9}
\end{equation}

From an explicit form for $C_{\mu \nu}^{\alpha \beta}$ (see 
Ref.~\cite{Kisselev-96}, Appendix~I, for details) one 
can verify that it obeys the same condition:
\begin{equation}
k^{\alpha} C_{\alpha \beta}^{\mu \nu} = 0. \label{10}
\end{equation}

Equations~(\ref{9}) and (\ref{10}) allow us to simplify expression~(\ref{4}) 
and get ($r=a,b$):
\begin{equation}
\frac{1}{x} F_2^{(r)} = \int \frac{d^4k}{(2\pi)^4} \frac{1}{k^4}
C_{\alpha \beta}^{(r)} (q,k;m_q) \Gamma^{\alpha \beta} (k,p), \label{11}
\end{equation}
with the notations
\begin{eqnarray}
C_{\alpha \beta}^{(a)} &=& - g_{\mu \nu} C_{\alpha \beta}^{\mu \nu},
\nonumber \\
C_{\alpha \beta}^{(b)} &=&  \frac{3Q^2}{(pq)^2 + Q^2 M^2} p_{\mu} p_{\nu} 
C_{\alpha \beta}^{\mu \nu}. \label{12}
\end{eqnarray}

The tensor $\Gamma^{\alpha \beta}$ can be expanded in Lorentz structures
\begin{eqnarray}
\Gamma^{\alpha \beta} &=& \biggl(g_{\alpha \beta} - \frac{k_{\alpha} 
k_{\beta}}{k^2} \biggl) \Gamma_1 + \biggl(p_{\alpha} - k_{\alpha}
\frac{pk}{k^2} \biggl)\biggl(p_{\beta} - k_{\beta} \frac{pk}{k^2} \biggl)
\frac{1}{k^2} \Gamma_2 \nonumber \\
&+& \biggl(k_{\alpha} - n_{\alpha} \frac{k^2}{kn} \biggl)
\biggl(k_{\beta} - n_{\beta} \frac{k^2}{kn} \biggl) \frac{1}{k^2} \Gamma_3 
+ \biggl(p_{\alpha} - n_{\alpha} \frac{pk}{kn} \biggl)
\biggl(p_{\beta} - n_{\beta} \frac{pk}{kn} \biggl) \frac{1}{k^2}
\Gamma_4 \label{13}
\end{eqnarray}
with $\Gamma_i=\Gamma_i (k^2,M^2,pk)$.

Let us consider a contribution of the invariant function $\Gamma_1$ into the 
structure function $F_2$~(\ref{11}). With the accounting for (\ref{9})
and (\ref{10}) we obtain
\begin{equation}
\frac{1}{x} F_2^{(r)} = e_q^2 \int \limits_x^1 \frac{dz}{z}
\int \limits_{Q_0^2}^{Q^2(z/x)} \frac{dl^2}{l^2}
\frac{1 - l^2 x^2/Q^2 z^2}{1 + M^2 x^2/Q^2} C^{(r)} \left( 
\frac{Q^2}{l^2}, \frac{m_q^2}{l^2}, \frac{x}{z} \right) 
\frac{\partial}{\partial \ln l^2} g(l^2,z), \label{14}
\end{equation}
where
\begin{equation}
l^2 = - k^2 > 0, \label{15}
\end{equation}
\begin{equation}
z = \frac{kn}{pn} \label{16}
\end{equation}
and
\begin{equation}
Q_0^2 = \frac{M^2 z^2}{1 - z}. \label{17}
\end{equation}
Here we used the notation:
\begin{equation}
C^{(r)} = - g^{\alpha \beta} C_{\alpha \beta}^{(r)}. \label{18}
\end{equation}

To be more correct, one has to write $z>x(1+4m^2/Q^2)$ and 
$l^2<Q(z/x) - 4m^2z/(z-x)$ in (\ref{14}), but me neglect power corrections
O($m^2/Q^2$).

In Eq.~(\ref{14}) the gluon distribution, $g(l^2,z)$, is introduced:
\begin{equation}
g(l^2,z) = \frac{1}{2(2\pi)^4} \int \limits_{Q_0^2}^{l^2} \frac{dl'^2}{l'^4}
\int d^2 k_{\bot} \Gamma_1 (l'^2,k_{\bot},z). \label{19}
\end{equation}
If we use the new variable
\begin{equation}
\xi = \frac{-k^2}{pk + \sqrt{(pk)^2 - k^2 M^2}} \label{20}
\end{equation}
instead of $k_{\bot}^2$, we will arrive at the expression
\begin{equation}
g(l^2,z) = \frac{z}{32\pi^3} \int \limits_{Q_0^2}^{l^2} \frac{dl'^2}{l'^4}
\int \limits_z^1 d\xi \left( M^2 + \frac{l'^2}{\xi^2} \right)
\Gamma_1 (l'^2,\xi). \label{21}
\end{equation}

A thorough analysis shows, however, that the main contribution to $F_2$ 
at small $x$ comes from $\Gamma_2$ and $\Gamma_4$ in~(\ref{13}) and
$F_2$ is given by the formula (see~\cite{Kisselev-96} for details):
\begin{eqnarray}
\frac{1}{x} F_2 = e_q^2 \sum_{r=a,b} \int \limits_z^1 \frac{dz}{z}
\int \limits_{Q_0^2}^{Q^2(z/x)} \frac{dl^2}{l^2} \!\!\! 
&& \biggl[ \tilde C^{(r)} \left( \frac{Q^2}{l^2}, \frac{m_q^2}{l^2}, 
\frac{x}{z} \right) \frac{\partial}{\partial \ln l^2} G(l^2,z) 
\nonumber \\
&& + \ \hat C^{(r)} \left( \frac{Q^2}{l^2}, \frac{m_q^2}{l^2}, 
\frac{x}{z} \right) \frac{\partial}{\partial \ln l^2} \hat G(l^2,z)  
\biggr]. \label{22}
\end{eqnarray}

As we are interested in a calculation of the difference of the structure 
functions corresponding to the massive and massless cases, we preserve
those terms in $C^{(r)}$ which give a leading contribution to $\Delta F_2$.
In~\cite{Kisselev-96} we have calculated the functions $C^{(a)}$ 
in lowest order in the strong coupling $\alpha_s$:
\begin{eqnarray}
\tilde C^{(a)} (u,v,y) &=&  \frac{\alpha_s}{4\pi} \{ [(1 - y)^2 + y^2] L(u,v,y) 
- [(1 - y)^2 + y^2 - 2v] M(v,y) - 1\}, \nonumber \\
\hat C^{(a)} (u,v,y) &=& \frac{\alpha_s}{\pi} y(1 - y) M(v,y) , \label{23}
\end{eqnarray}
where
\begin{eqnarray}
L(u,v,y) &=& \ln \frac{u(1-y)}{y [v + y(1 - y)]} \raise 2pt \hbox{,}
\nonumber \\
M(v,y) &=& \frac{y(1-y)}{v + y(1 - y)} \raise 2pt \hbox{.} \label{24}
\end{eqnarray}

As for the gluon distributions, they are given by the formulae:
\begin{eqnarray}
G &=& \frac{1}{32\pi^3z} \int \limits_{Q_0^2}^{l^2} \frac{dl'^2}{l'^4}
\int \limits_z^1 \frac{d\xi}{\xi} (\xi - z) \left( M^2 + \frac{l'^2}{\xi^2} 
\right)
[\Gamma_2 (l'^2,\xi) + \Gamma_4 (l'^2,\xi)], \label{25a} \\
\hat G &=& \frac{1}{32\pi^3z} \int \limits_{Q_0^2}^{l^2} \frac{dl'^2}{l'^4}
\int \limits_z^1 d\xi 
\left( M^2 + \frac{l'^2}{\xi^2} \right) 
\left[ \frac{(2\xi - z)^2}{4\xi^2} 
\Gamma_2 (l'^2,\xi) + \Gamma_4 (l'^2,\xi) \right]. \label{25b}
\end{eqnarray}

The analogous expressions for the functions $C^{(b)}$ are 
the following~\cite{Kisselev-96}:
\begin{eqnarray}
\tilde C^{(b)} (u,v,y) &=& \frac{3\alpha_s}{2\pi}
\frac{1}{u} y \{ 2y [(1 - 2y) (1 - y) -v] L(u,v,y) \nonumber \\
&+& (1 - y) [(1 - y)^2 + y^2 - 2v] M(v,y) \} 
+ \frac{3\alpha}{2\pi} y(1 - y), \nonumber \\
\hat C^{(b)} (u,v,y) &=& - \frac{12\alpha_s}{\pi}
\frac{1}{u} y^2 (1 - y)^2 M(v,y). \label{26}
\end{eqnarray}

It may be shown that the leading contribution to $\Delta F_2$ comes from the 
region $l^2 \sim m_q^2$, $k^2=-l^2$ being the gluon virtuality.
Then one can easily see from (\ref{24}) and (\ref{26}) that the first two 
terms in 
$\tilde C^{(b)}$ are suppressed  by the factor $k^2/Q^2$ with respect to 
$\tilde C^{(a)}$,
while the third terms in $\tilde C^{(b)}$ do not contribute to the difference 
$C^{(b)}|_{m = 0} - C^{(b)}|_{m \neq 0}$. 

In the leading logarithmic approximation (LLA), only the function $L$  
remaines in Eqs.~(\ref{23}), which results in
\begin{equation}
\frac{1}{x} \frac{\partial}{\partial \ln Q^2} F_2(Q^2,x) = 
\frac{\alpha_s}{2\pi} \int \limits_x^1 \frac{dz}{z}P_{qg} \biggl( \frac{x}{z}
\biggr) G(Q^2,z) \raise 2pt \hbox{,} \label{27}
\end{equation}
where $P_{qg}(z)$ is the Altarelli--Parisi splitting function and $G(Q^2,z)$ 
is the gluon distribution in LLA defined by Eq.~(\ref{25a}). 

It is clear from (\ref{22}) that 
$\Delta F_2 = F_2|_{m = 0} - F_2|_{m \neq 0}$ is defined by the 
quantities ($r=a,b$)
\begin{equation}
\Delta C^{(r)}(u,v,y) = C^{(r)}(u,0,y) - C^{(r)}(u,v,y). \label{30}
\end{equation}
By using Eq.~(\ref{23}) we obtain the important result
\begin{eqnarray}
\Delta \tilde C^{(a)} &=& \Delta \tilde C^{(a)}(v,y) \nonumber \\
\Delta \hat C^{(a)} &=& \Delta \hat C^{(a)}(v,y), \label{31}
\end{eqnarray}
while from (\ref{26}) we get
\begin{eqnarray}
\Delta \tilde C^{(b)} &=& \frac{1}{u} \Delta \tilde C^{(b)}(v,y),
\nonumber \\
\Delta \hat C^{(b)} &=& \frac{1}{u} \Delta \hat C^{(b)}(v,y). \label{32}
\end{eqnarray}
In this, we have
\begin{equation}
\Delta \tilde C^{(a)}, \Delta \hat C^{(a)} 
|\mathstrut_{-k^2 \rightarrow \infty} \sim \frac{m_q^2}{k^2}. 
\label{33}
\end{equation}
So, we get~\cite{Kisselev-96}
\begin{eqnarray}
\frac{1}{x} \Delta F_2 (Q^2,m_q^2,x) |\mathstrut_{Q^2 \rightarrow 
\infty} = e_q^2 \int \limits_x^1 \frac{dz}{z}
\int \limits_{Q_0^2}^{\infty} \frac{dl^2}{l^2} \!\!\! 
&& \biggl[ \Delta \tilde C \left( \frac{m_q^2}{l^2},\frac{x}{z} \right) 
\frac{\partial}{\partial \ln l^2} G(l^2,z) \nonumber \\ 
&& + \ \Delta \hat C \left( \frac{m_q^2}{l^2}, \frac{x}{z} \right)
\frac{\partial}{\partial \ln l^2} \hat G(l^2,z) \biggr]. \label{34}
\end{eqnarray}
Here
\begin{eqnarray}
\Delta \tilde C(v,y) &=& \frac{\alpha_s}{4\pi} \biggl\{ [(1 - y)^2 + y^2] 
\ln \biggl[ 1 + {\displaystyle \frac{v}{y(1-y)}} \biggr] 
- {\displaystyle \frac{v}{v + y(1-y)}} \biggr\}, \nonumber \\
\Delta \hat C(v,y) &=& \frac{\alpha_s}{\pi} 
y(1-y) {\displaystyle \frac{v}{v + y(1-y)}} \label{35}
\end{eqnarray}
with $G(l^2,z)$ and $\hat G(l^2,z)$ being defined by Eqs.~(\ref{25a})
and (\ref{25b}). 

The integral in $l^2$ (\ref{34}) converges because of condition~({\ref{33}). 
Contributions from $\Delta \tilde C^{(b)}$ and $\Delta \hat C^{(b)}$ are 
suppressed by the factors $(m^2/Q^2) \ln Q^2$ and can thus be omitted.

Let us consider the gluon distribution $\hat G$~(\ref{25b}). At small $z$ the
leading contribution to $\hat G(l^2,z)$ comes from the region $z \ll \xi$,
and we have
\begin{equation}
\hat G(l^2,z) \simeq G(l^2,z). \label{36}
\end{equation}
Taking expression~(\ref{36}) into account, the structure function 
$F_2$~(\ref{22}) has the following form at low $x$ (with the term of the order
of $k^2/Q^2$ and $m^2/Q^2$ subtracted)
\begin{equation}
\frac{1}{x} F_2 = e_q^2 \int \limits_x^1 \frac{dz}{z} 
\int \limits_{Q_0^2}^{Q^2(z/x)} \frac{dl^2}{l^2} C \left( \frac{Q^2}{l^2},
\frac{m_q^2}{l^2}, \frac{x}{z} \right) 
\frac{\partial}{\partial \ln l^2} G(l^2,z), \label{37}
\end{equation}
where
\begin{equation}
C(u,v,y) = \frac{\alpha_s}{4\pi} \{ [(1 - y)^2 + y^2] L(u,v,y)
- [(1 - 3y)^2 - 3y^2 - 2v] M(v,y) - 1 \}. \label{38}
\end{equation}

As for the difference of the structure function, we obtain the following
prediction
\begin{equation}
\frac{1}{x} \Delta F_2 (Q^2, m_q^2, x) |\mathstrut_{Q^2 \rightarrow 
\infty}
= \frac{1}{x} \Delta F_2 (m_q^2,x) = e_q^2 \int \limits_x^1 \frac{dz}{z} 
\int \limits_{Q_0^2}^{\infty}
\frac{dl^2}{l^2} \Delta C \left( \frac{m_q^2}{l^2}, \frac{x}{z} \right)
\frac{\partial}{\partial \ln l^2} G(l^2,z), \label{39}
\end{equation}
where
\begin{equation}
\Delta C(v,y) = \frac{\alpha_s}{4\pi} [(1 - y)^2 + y^2] 
\biggl\{ \ln \biggl[1 + {\displaystyle \frac{v}{y(1 - y)}} \biggr] 
- (1 - 2y)^2 {\displaystyle \frac{v}{v + y(1 - y)}} \biggr\}. \label{40}
\end{equation}

\section{Relation between measurable structure functions}

Up to now, we considered those contributions to $F_2$ that came from the
quark with electric charge $e_q$ and mass $m_q$, $\tilde F_2|_{m \neq 0}$.
Then we have taken the analogous contributions from the massless quark with
the same $e_q$, $\tilde F_2|_{m = 0}$, and have calculated the quantity
$\tilde F_2|_{m = 0} - \tilde F_2|_{m \neq 0}$.

The total structure function $F_2$ has the form
\begin{equation}
F_2(Q^2,x) = \sum_q e_q^2 \tilde F_2^q(Q^2,x), \label{41}
\end{equation}
where the functions $\tilde F_2^q$ are introduced ($q = u,d,s,c,b$).

The structure functions describing open charm and bottom production in DIS,
$F_2^c$ and $F_2^b$ respectively, are related to $\tilde F_2^c$ and 
$\tilde F_2^b$ by the formulae
\begin{eqnarray}
F_2^c &=& \frac{4}{9} \tilde F_2^c, \nonumber \\
F_2^b &=& \frac{1}{9} \tilde F_2^b. \label{42}
\end{eqnarray}

At low $x$ one can put ($m_u = m_d = m_s = 0$ is assumed)
\begin{equation}
\tilde F_2^u = \tilde F_2^d = \tilde F_2^s = \tilde F_2 \label{43}
\end{equation}
and define the difference between heavy and light flavour contributions to
$F_2$:
\begin{eqnarray}
\Delta \tilde F_2^c = \tilde F_2 - \tilde F_2^c, \nonumber \\
\Delta \tilde F_2^b = \tilde F_2 - \tilde F_2^b. \label{44}
\end{eqnarray}
Notice that there are the functions $\tilde F_2$ and $\tilde F_2^q$ that 
have been calculated in the previous section (see Eqs~(\ref{37}) 
and (\ref{39})).

From Eqs.~(\ref{39}) and ({\ref{44}) one readily obtains that a linear
combination
\begin{equation}
\Sigma_\alpha (Q^2,x) \equiv F_2 (Q^2,x) + \alpha F_2^c (Q^2,x)
- (4 \alpha + 11) F_2^b (Q^2,x) \label{44.1}
\end{equation}
scales at $Q^2 \rightarrow \infty$ and arbitrary parameter $\alpha$. In
terms of $\Delta \tilde F_2$ introduced in~(\ref{39})
\begin{equation}
\lim_{Q^2 \rightarrow \infty} \Sigma_\alpha (Q^2,x) = -\frac{4}{9}
(1 + \alpha) \Delta \tilde F_2 (m_c^2,x) + \frac{1}{9} (4\alpha + 10)
\Delta \tilde F_2 (m_b^2,x). \label{44.2}
\end{equation}

Let us now represent the function $\tilde F_2$~(\ref{37}) in the following
form 
\begin{equation}
\frac{1}{x} \tilde F_2 = \int \limits_x^1 \frac{dy}{y} \int \limits_0^Y
d \eta \, C(\eta, y) G' \left(Y - \eta, 
\frac{x}{y} \right), \label{45}
\end{equation}
where we denote
\begin{equation}
Y = \ln \frac{Q^2}{y Q_0^2} \label{46}
\end{equation}
and introduce the variable $\eta = \ln (k^2/Q_0^2)$. Here $G'$ means the 
derivative of $G(Q^2,x)$ with respect to the variable $\ln Q^2$.

Analogously, we get from (\ref{39})
\begin{equation}
\frac{1}{x} \Delta \tilde F_2^q = \int \limits_x^1 \frac{dy}{y} \int 
\limits_{- \infty}^{Y_m} d \eta \, \Delta C(\eta, y) 
G' \left( Y_m - \eta, \frac{x}{y} \right), \label{47}
\end{equation}
with
\begin{equation}
Y_m = \ln \frac{m_q^2}{y Q_0^2}. \label{48}
\end{equation}
Here $\eta = \ln (m_q^2/k^2 y(1-y)) \simeq \ln (m_q^2/k^2 y)$ (remember that
we consider small $x$).

The expression for $\Delta C$ is given by Eq.~(\ref{40}) and,  
in terms of the variables $\eta$ and $y$, looks like 
\begin{equation}
\Delta C = \frac{\alpha_s}{4\pi} [(1 - y)^2 + y^2] \biggl[ \ln \left(
1 + e^{\eta} \right) - (1 - 2y)^2 \frac{e^{\eta}}{1 + e^{\eta}} \biggr].
\label{49}
\end{equation}

As for the expression for $C$, it has to be defined via relation~(\ref{11})
and exact formulae (\cite{Kisselev-96}) taken at $m = 0$. The result
is of the form
\begin{equation}
C(\eta, y) = \frac{\alpha_s}{2\pi} \left[ \frac{1}{2U} \ln \frac{1 + U}{1 - U}
\biggl( 1 - \frac{3}{U^2} V + V \biggr) - \biggl( 1 - \frac{3}{U^2} V \biggr)
\right], \label{50}
\end{equation}
where
\begin{eqnarray}
U &=& \sqrt{1 - 4y (1 - y) e^{-\eta}}, \nonumber \\
V &=& (1 - y) \left[ y + (1 - y) e^{-\eta} \right] 
\left( 1 - e^{-\eta} \right). \label{51}
\end{eqnarray}

It is clear from (\ref{49}) that
\begin{equation}
\Delta C(\eta, y) > 0 \label{52}
\end{equation}
for $- \infty < \eta < \infty$, $0 \leq y \leq 1$ and $\Delta C(\eta, y)$
is negligible at $\eta < 0$ (see Figs.~1a-1d).

Moreover, the quantitative analysis shows that at most at $y \leq 0.2$, 
which is relevant for small $x$ as under consideration, one has
\begin{equation}
C(\eta, y) > \Delta C(\eta, y), \qquad \eta > 0, \label{53}
\end{equation}
(see Figs.~2a-2d).
Neglecting the small contribution to $\tilde F_2$ from the region $\eta < 0$
and taking into account that $\partial G(Q^2, x) / \partial \ln Q^2 > 0$ at
small $x$ (cf.~\cite{Kwiecinski-96}), we thus conclude
\begin{equation}
\Delta \tilde F_2^q (m_q^2, x) < \tilde F_2 (Q^2, x)|_{Q^2 = m_q^2}. \label{54}
\end{equation}

From Eqs.~(\ref{44.2}), (\ref{54}) we obtain the following inequality which
holds for $-2.5 \leq \alpha \leq -1$
\begin{eqnarray}
0 < \Sigma_\alpha (Q^2,x) |\mathstrut_{Q^2 \gg 1{\rm GeV}^2}
&<& - \frac{2}{3} (1 + \alpha) ( F_2 - F_2^c - F_2^b) (Q^2,x)|_{Q^2 = m_c^2}
\nonumber \\
&& +\frac{1}{3} (5 + 2\alpha) ( F_2 - F_2^c - F_2^b) (Q^2,x)|_{Q^2 = m_b^2}.
\label{55}
\end{eqnarray}
At the endpoints $\alpha=-2.5$ and $\alpha=-1$ we get
\begin{eqnarray}
\left( F_2 - 2.5 F_2^c - F_2^b \right) (Q^2,x) &<& 
\left( F_2 - F_2^c - F_2^b \right) (Q^2,x) |_{Q^2 = m_c^2},
\nonumber \\
\left( F_2 - F_2^c -7 F_2^b \right) (Q^2,x) &<& 
\left( F_2 - F_2^c - F_2^b \right) (Q^2,x)|_{Q^2 = m_b^2}. \label{56}
\end{eqnarray}

Data on the total structure function $F_2$ for $Q^2$ between $1.5 \ 
{\rm GeV}^2$ and $5000 \ {\rm GeV}^2$ and $x$ between $3 \times 10^{-5}$
and $0.32$ are now available~\cite{H1-96}. As for the charm structure 
function, there are recent data on $F_2^c$ at $Q^2 = 12 \ 
{\rm GeV}^2$, $25 \ {\rm GeV}^2$ and $45 \ {\rm GeV}^2$ with rather large
errors~\cite{H1-96*}. 

Using the first of the inequalities~(\ref{56}) we get (assuming 
$F_2^c (m_c^2, x)$,  $F_2^b (m_c^2, x) \simeq 0$ (cf.~\cite{Martin-94}))
\begin{equation}
F_2^c (Q^2, x) > 0.4 \left[ F_2(Q^2, x) - F_2^b (Q^2, x) - F_2 (m_c^2, x)
\right]. \label{57}
\end{equation}

Taking use of the available data on $F_2(Q^2,x)$~\cite{H1-96} and neglecting
$F_2^b$ (as $F_2^b/F_2$ reaches at most $2 \div 3 \%$ at HERA) we estimate 
the lower bound according to~(\ref{57}). The result is exhibited in 
Figs.~3a-3c with $m_c^2 = 2.5$ GeV$^2$. Experimental data on $F_2^c$ at 
several values of $Q^2$ and $x$ are taken from Ref.~\cite{H1-96*}.

Tables~1-3 present the result of our calculations of the lower bounds on
$F_2^c$ at the same $Q^2$ and $x$. Three different values of $F_2^c$ for
each $Q^2$, $x$ correspond to $m_c = 1.3$ GeV, $m_c = 1.5$ GeV,
$m_c = 1.7$ GeV, respectively.
\begin{center}
\begin{tabular}{||c|c|c||}
\hline
$< x >$ & $F_2^c$ (theor.) & $F_2^c$ (exper.) \cite{H1-96*} \\
\hline 
       & 0.173 & \\
.0008  & 0.161 & $0.211 \pm 0.049\ ^{+0.045}_{-0.040}$ \\
       & 0.145 & \\
\hline
       & 0.137 & \\
.0020  & 0.128 & $0.263 \pm 0.036\ ^{+0.043}_{-0.041}$ \\
       & 0.116 & \\ 
\hline
       & 0.120 & \\
.0032  & 0.112 & $0.190 \pm 0.054\ ^{+0.052}_{-0.049}$ \\
       & 0.101 & \\
\hline
\end{tabular}\\
\vspace*{.3cm}
Table~1. The lower bounds on $F_2^c(Q^2,x)$ for $Q^2=12$ GeV$^2$. \\
\vspace*{.7cm}
\begin{tabular}{||c|c|c||}
\hline
$< x >$ & $F_2^c$ (theor.) & $F_2^c$ (exper.) \cite{H1-96*} \\
\hline 
       & 0.258 & \\
.0008  & 0.247 & $0.324 \pm 0.099\ ^{+0.065}_{-0.058}$ \\
       & 0.231 & \\
\hline
       & 0.205 & \\
.0020  & 0.196 & $0.253 \pm 0.069\ ^{+0.041}_{-0.040}$ \\
       & 0.184 & \\ 
\hline
       & 0.179 & \\
.0032  & 0.172 & $0.222 \pm 0.066\ ^{+0.044}_{-0.039}$ \\
       & 0.161 & \\
\hline
\end{tabular}\\
\vspace*{.3cm}
Table~2. The lower bounds on $F_2^c(Q^2,x)$ for $Q^2=25$ GeV$^2$. \\
\vspace*{.7cm}
\begin{tabular}{||c|c|c||}
\hline
$< x >$ & $F_2^c$ (theor.) & $F_2^c$ (exper.) \cite{H1-96*} \\ 
\hline 
       & 0.258 & \\
.0020  & 0.249 & $0.156 \pm 0.070\ ^{+0.031}_{-0.028}$ \\
       & 0.237 & \\
\hline
       & 0.226 & \\
.0032  & 0.218 & $0.275 \pm 0.074\ ^{+0.045}_{-0.042}$ \\
       & 0.207 & \\ 
\hline
       & 0.165 & \\
.0080  & 0.160 & $0.200 \pm 0.064\ ^{+0.040}_{-0.035}$ \\
       & 0.152 & \\
\hline
\end{tabular}\\
\vspace*{.3cm}
Table~3. The lower bounds on $F_2^c(Q^2,x)$ for $Q^2=45$ GeV$^2$. \\
\end{center}

These estimates of $F_2^c$ agree with the recent
data on the charm contribution to $F_2$~\cite{H1-96*}.
Our inequalities~(\ref{55})-(\ref{57}) are also in agreement with
the results of Ref.~\cite{Laenen-92} where the ratio $F_2^c/F_2$ was 
estimated. 
For a detailed comparison of our predictions with the data, an improved 
measurement of the charm component $F_2^c$ is required.

\section*{Conclusions}

In this paper we have demonstrated that the lowest--order quark loop 
contributions to the structure functions at small $x$ contain mass--dependent
terms which scale at high $Q^2$. This effect can be observed experimentally,
and we predict theoretical bounds for the corresponding contributions from
$c$-quarks (see Eqs.~(\ref{55}) and (\ref{56}), Figs.~3a-3c, 
Tabs.~1-3).

\section*{Acknowledgements}

We are grateful to Profs. G. Veneziano and J.B. Dainton and to Dr. E. Laenen 
for stimulating conversations. 
We also thank Drs. A. De Roeck, M. Kuhlen and U. Karshon for a useful 
discussion of the data on $F_2^c$ and for sending us the recent data 
on $F_2$ from the H1 Collaboration.
\vfill \eject

\vfill \eject

\section*{Figure Captions}

\noindent
{\bf Figs. 1a-1d:} \ $\Delta C(\eta, y)$ as a function of the variable $\eta$  
at several fixed values of $y$. \\

\noindent
{\bf Figs. 2a-2d:} \ $C(\eta, y)$ (continuous curves) and $\Delta C(\eta, y)$ 
                    (dashed curves) as functions of the \\ 
\hphantom{\bf Figs. 2a-2d:} \ 
                    variable $\eta$ ($\eta \geq 0$) at several fixed   
                    values of $y$.  \\

\noindent
{\bf Figs. 3a-3c:} \ The lower bounds on $F_2^c(Q^2,x)$ (cotinuous curves)
                    together with results \\
\hphantom{\bf Figs. 3a-3c:} \
                    from H1~\cite{H1-96*} (open and closed circles).
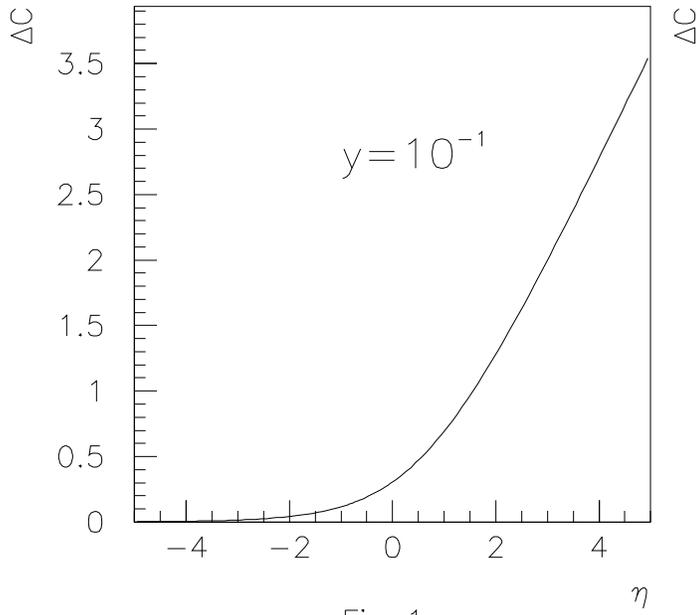
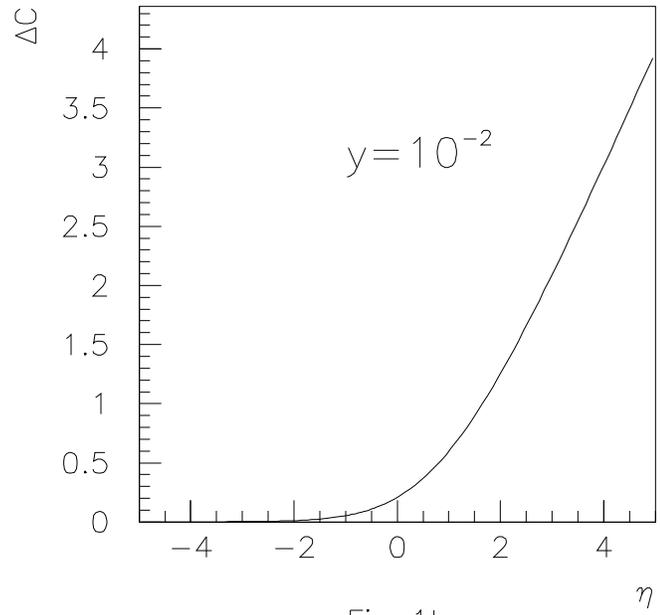
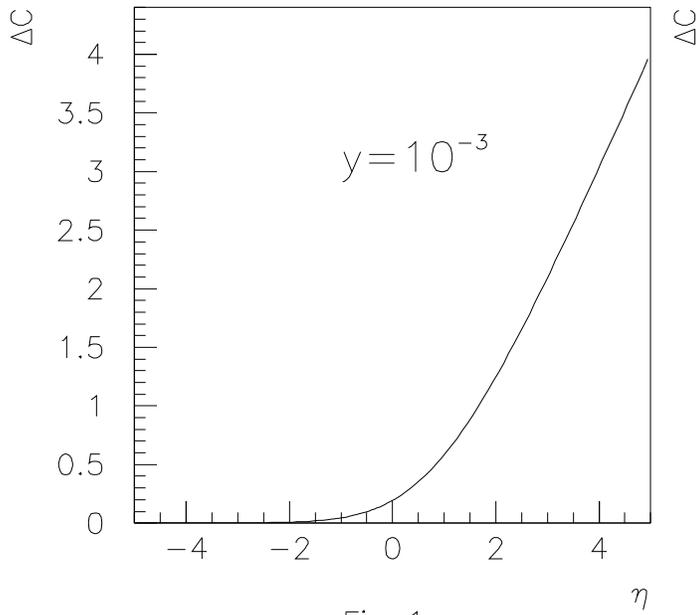
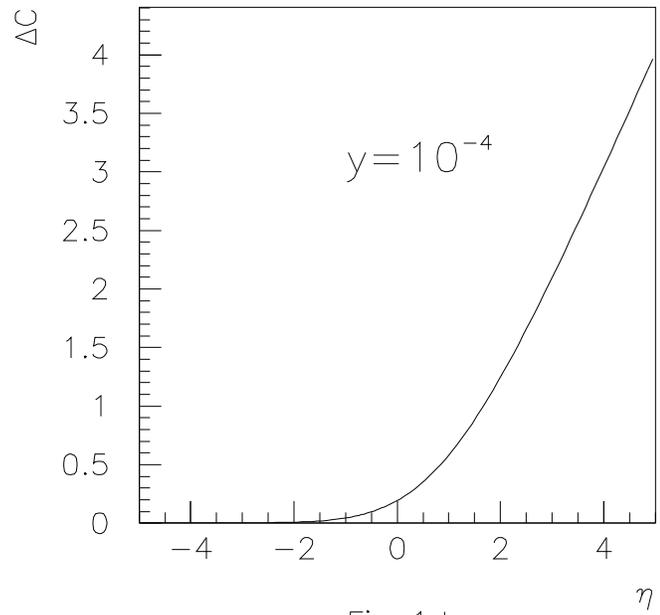
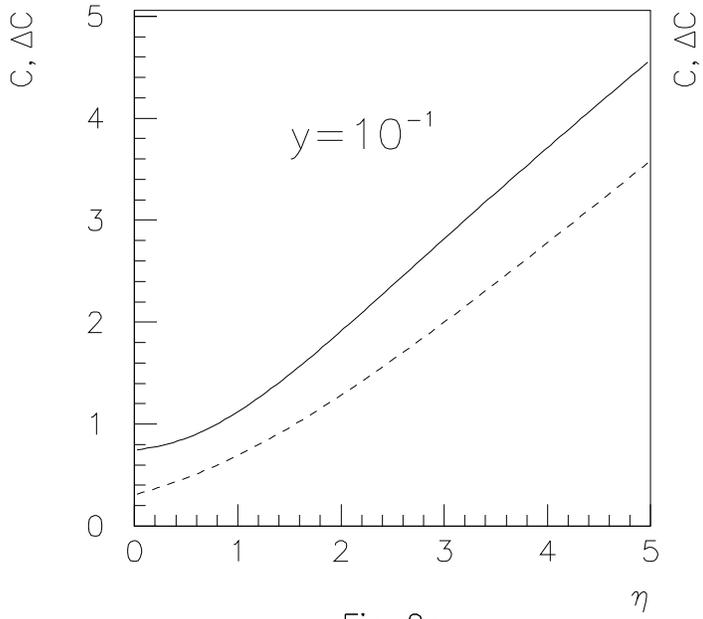
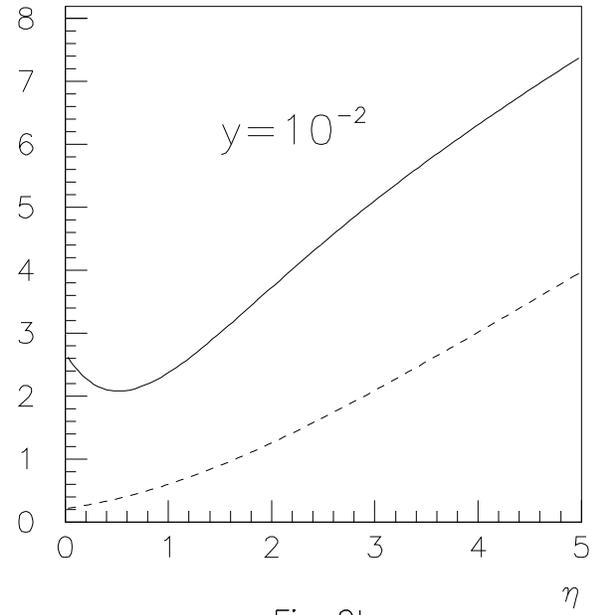
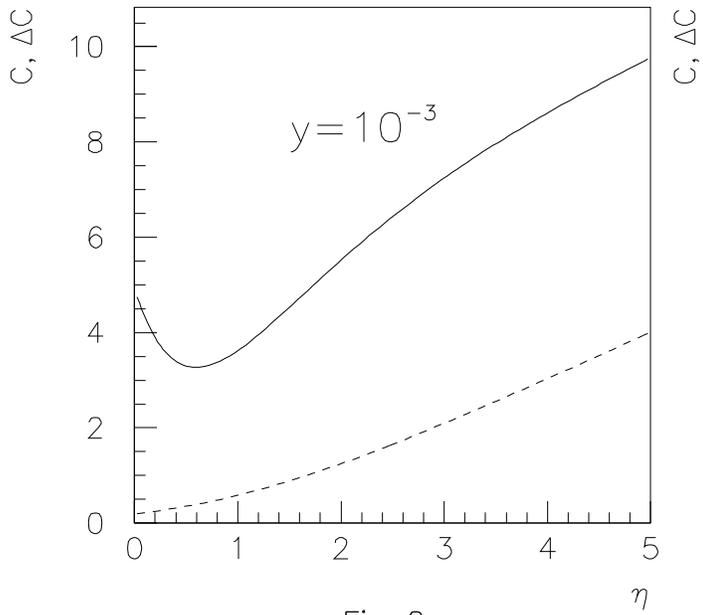
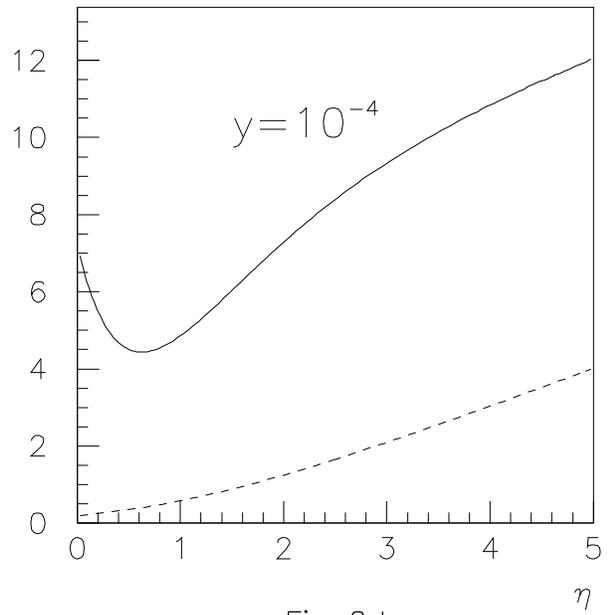
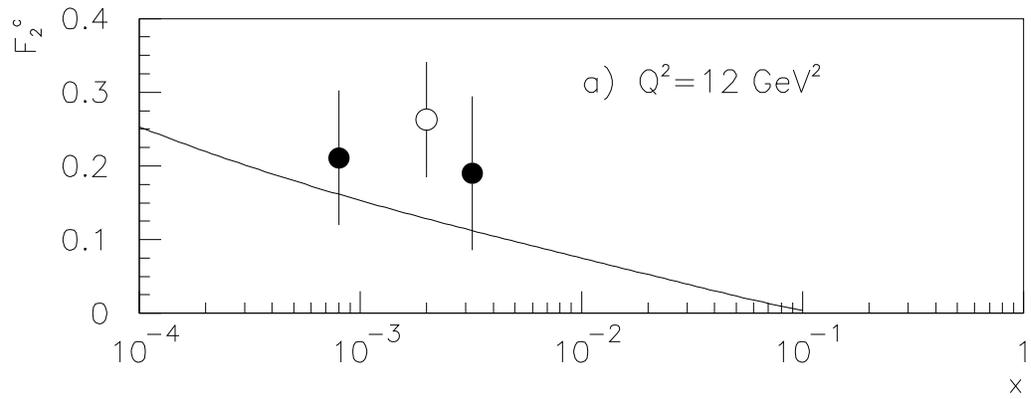
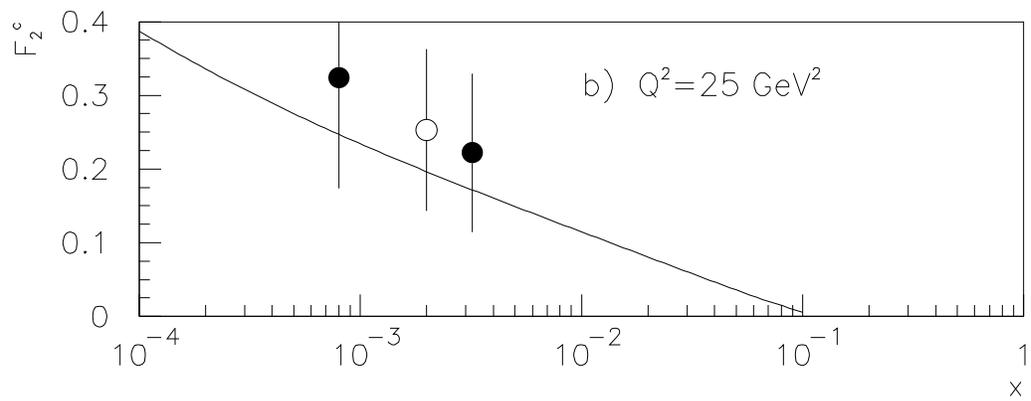
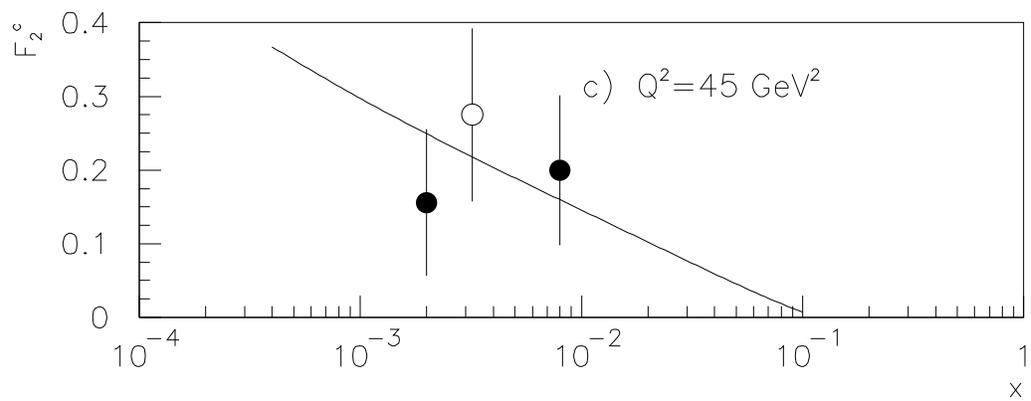
\end{document}